\documentclass[acmsmall]{acmart}
\usepackage{etoolbox}
\makeatletter
\patchcmd{\@maketitle}{\@thanks}{}{}{}
\makeatother
\AtBeginDocument{%
  }

\setcopyright{cc}
\setcctype{by}
\acmJournal{PACMMOD}
\acmYear{2025} \acmVolume{3} \acmNumber{6 (SIGMOD)} \acmArticle{316} \acmMonth{12} \acmPrice{}\acmDOI{10.1145/3769781}

\usepackage{subcaption}
\usepackage{svg}
\usepackage{lineno}
\usepackage{makecell}
\usepackage{multirow}
\usepackage{listings}
\usepackage{balance}
\usepackage{tabularx}
\usepackage{array}
\usepackage{xspace}
\usepackage[edges]{forest}
\usepackage{tikz-cd}
\usepackage{tikz}
\usepackage{threeparttable}
\usetikzlibrary{fit, positioning, backgrounds}
\usepackage{booktabs}
\usepackage{pifont}
\usepackage{fancyvrb}
\usepackage[most]{tcolorbox} 
\tcbuselibrary{listingsutf8}

\newcommand{\paradigmname}{\textsc{Drama}\xspace}
\newcommand{\sysname}{\textsc{DramaBot}\xspace}
\newcommand{\datasetname}{\textsc{DramaBench}\xspace}
\newcommand{\minihead}[1]
{{\vspace{.5em}\noindent\textbf{#1.} }}

\definecolor{asterick-color}{HTML}{fdb863}
\definecolor{open-color}{HTML}{D5E8D4}
\definecolor{darkgreen}{HTML}{33a02c}
\definecolor{darkred}{HTML}{e31a1c}
\definecolor{lightgraybox}{HTML}{F5F5F5}
\definecolor{grayborder}{HTML}{C0C0C0}
\lstdefinelanguage{CustomSQL}{
  basicstyle=\ttfamily\small,
  keywordstyle=\color{blue},
  morekeywords={SELECT, FROM, WHERE, GROUP, BY, COUNT, AS, LIKE, ORDER, DESC, LIMIT, JOIN},
  sensitive=true,
  commentstyle=\color{gray},
  stringstyle=\color{red},
  showstringspaces=false,
    literate={
    {1}{{{\color{teal}1}}}1
    {\%NP}{{{\color{teal}\%NP}}}2
  },
}

\definecolor{commentgray}{gray}{0.45}
\definecolor{cmdblue}{rgb}{0.0,0.2,0.6}
\definecolor{earthyellow}{rgb}{0.72, 0.53, 0.04}
\lstdefinestyle{cmdstyle}{
  language=bash,
  basicstyle=\ttfamily\small,
  breaklines=true,
  keywordstyle=\color{cmdblue},
  commentstyle=\color{commentgray}\itshape,
  morekeywords={curl,echo,mv,unzip},
  emph={[1]{>, -OJ, --output-dir, *}},
  emphstyle={[1]\color{teal}},
  emph={[2]{$R}},
  emphstyle={[2]\color{earthyellow}},
  alsoletter={-, *}>$,
  showstringspaces=false,
}

\newtcblisting{cmdbox}{
  listing only,
  colback=gray!5,
  colframe=gray!50,
  listing options={
    basicstyle=\ttfamily\small,
    breaklines=true,
    keywordstyle=\color{cmdblue},
    commentstyle=\color{commentgray}\itshape,
    morekeywords={curl,echo,mv,unzip},
    alsoletter={<>=},
    showstringspaces=false,
  },
  left=2mm, right=2mm, top=0.05mm, bottom=0.05mm,
  enhanced,
  boxrule=0.4pt,
}

\lstdefinelanguage{CustomSQL1}{
  basicstyle=\ttfamily\small,
  keywordstyle=\color{blue},
  morekeywords={SELECT, FROM, WHERE, GROUP, BY, COUNT, AS, LIKE, ORDER, DESC, LIMIT, JOIN, ON},
  sensitive=true,
  commentstyle=\color{gray},
  stringstyle=\color{red},
  showstringspaces=false,
  literate={*}{{{\color{teal}*}}}1,
}

\usepackage{xcolor}

\begin{document}

\title[\paradigmname: Unifying Data Retrieval and Analysis for Open-Domain Analytic Queries]{\paradigmname: Unifying Data Retrieval and Analysis \\for Open-Domain Analytic Queries}

\author{Chuxuan Hu}
\affiliation{
  \institution{University of Illinois Urbana-Champaign}
  \city{Urbana}
  \state{Illinois}
  \country{USA}
}
\email{chuxuan3@illinois.edu}

\author{Maxwell Yang}
\affiliation{
  \institution{University of Illinois Urbana-Champaign}
  \city{Urbana}
  \state{Illinois}
  \country{USA}
}

\author{James Weiland}
\affiliation{
  \institution{University of Illinois Urbana-Champaign}
  \city{Urbana}
  \state{Illinois}
  \country{USA}
}

\author{Yeji Lim}
\affiliation{
  \institution{University of Illinois Urbana-Champaign}
  \city{Urbana}
  \state{Illinois}
  \country{USA}
}

\author{Suhas Palawala}
\affiliation{
  \institution{University of Illinois Urbana-Champaign}
  \city{Urbana}
  \state{Illinois}
  \country{USA}
}

\author{Daniel Kang}
\affiliation{
  \institution{University of Illinois Urbana-Champaign}
  \city{Urbana}
  \state{Illinois}
  \country{USA}
}
\email{ddkang@illinois.edu}

\renewcommand{\shortauthors}{Chuxuan Hu et al.}

\begin{abstract}

Manually conducting real-world data analyses is labor-intensive and inefficient. Despite numerous attempts to automate data science workflows, none of the existing paradigms or systems fully demonstrate all three key capabilities required to support them effectively: (1) open-domain data collection, (2) structured data transformation, and (3) analytic reasoning.

To overcome these limitations, we propose \paradigmname, an end-to-end paradigm that answers users' analytic queries in natural language on large-scale open-domain data. \paradigmname unifies data collection, transformation, and analysis as a single pipeline.
To quantitatively evaluate system performance on tasks representative of \paradigmname, we construct a benchmark, \datasetname, consisting of two categories of tasks: claim verification and question answering, each comprising 100 instances. These tasks are derived from real-world applications that have gained significant public attention and require the retrieval and analysis of open-domain data.
We develop \sysname, a multi-agent system designed following \paradigmname. It comprises a data retriever that collects and transforms data by coordinating the execution of sub-agents, and a data analyzer that performs structured reasoning over the retrieved data. We evaluate \sysname on \datasetname together with five state-of-the-art baseline agents. \sysname achieves 86.5\% task accuracy at a cost of \$0.05, outperforming all baselines with up to $6.9\times$ the accuracy and less than 1/6 of the cost. \paradigmname is publicly available at \url{https://github.com/uiuc-kang-lab/drama}.

\end{abstract}



\received{April 2025}
\received[revised]{July 2025}
\received[accepted]{August 2025}

\maketitle

\section{Introduction}
\label{sec:intro}

\begin{figure}[t]
\centering
\resizebox{0.95\columnwidth}{!}{
\begin{tikzpicture}[
    font=\sffamily,
    stage/.style={
         draw=grayborder, fill=lightgraybox, rounded corners=8pt,
        minimum width=2.3cm, minimum height=1.6cm,
        text width=1.6cm, align=center
    },
    groupbox/.style={
        draw, rounded corners=12pt, inner sep=0.4cm
    },
    arrow/.style={->, thick}
]

\node[stage, text width=7.5cm, minimum width=6.5cm] (retrieving) at (0,0) {\textbf{\textcolor{blue}{(C1)} Data Collection}\\Open-Domain IE Systems \cite{10.1145/988672.988687, 10.5555/2283396.2283398, kamp2023openinformationextractionreview} \\ Web Search Tools \& Agents \cite{openai2024tools, he2024webvoyagerbuildingendtoendweb, btahir2024open, openai2024researchbot, yang2023autogptonlinedecisionmaking}};
\node[stage, text width=7.5cm, minimum width=6cm, below=0.55cm of retrieving] (management) {\textbf{\textcolor{blue}{(C2)} Database Construction}\\ Knowledge Base Construction \cite{liška2022streamingqabenchmarkadaptationnew, wang2023knowledgptenhancinglargelanguage, kasai2024realtimeqawhatsanswer} \\RAG-based Systems \cite{biswal2024text2sqlenoughunifyingai, Chen_2023, tang2024minicheckefficientfactcheckingllms}};
\node[stage, text width=5cm, minimum width=5cm, right=1.5cm of retrieving] (preproc) 
  {\textbf{Data Preprocessing}\\ Data Labeling \cite{Hu_2024}\\Data Cleaning \cite{cao2024spider2vfarmultimodalagents, hong2024datainterpreterllmagent}};
\node[stage, text width=5cm, minimum width=5cm, below=0.3cm of preproc] (query) 
  {\textbf{Semantic Querying}\\Query Generation \cite{biswal2024text2sqlenoughunifyingai, zhang2024datacopilotbridgingbillionsdata, dail_sql} \\ Query Execution \cite{patel2025semanticoperatorsdeclarativemodel, liu2024declarativeoptimizingaiworkloads, shankar2025docetlagenticqueryrewriting}\\ Question Answering \cite{eisenschlos-etal-2020-understanding}};

\draw[arrow] (retrieving.south) -- (management.north);
\draw[arrow] (preproc.south) -- (query.north);

\begin{pgfonlayer}{background}
\node[groupbox, fill=orange!20, draw=orange!70!black,
      fit=(retrieving)(query),
      inner ysep=1cm, text width=15cm, yshift=-0.01cm] (outerbox){};
\node at ([yshift=-0.3cm]outerbox.north) {\textbf{\paradigmname}};
\end{pgfonlayer}

\node[groupbox, draw=green!50!black, fill=none, dashed, very thick,
      fit=(retrieving)(management), inner ysep=0.5cm, yshift=-0.13cm, xshift=0.1cm] (collectbox) {};
\node at ([yshift=-4.4cm]collectbox.north) {\textbf{Data Retrieval}};

\node[groupbox, draw=yellow!50!black, fill=none, dashed, very thick,
      fit=(preproc)(query), inner ysep=0.5cm, yshift=-0.15cm, xshift=-0.1cm] (analysisbox) {};
\node at ([yshift=-4.4cm]analysisbox.north) {\textbf{\textcolor{blue}{(C3)} Data Analysis}};
\draw[arrow] (collectbox.east) -- (analysisbox.west);
\end{tikzpicture}%
}
\caption{\paradigmname integrates the full data science pipeline, where (C1)–(C3) correspond to the essential capabilities underpinning each stage. Here, ``Data Retrieval'' refers to the collection and structuring of raw data from open domains.}
\label{fig:dynamicdb-architecture}
\end{figure}
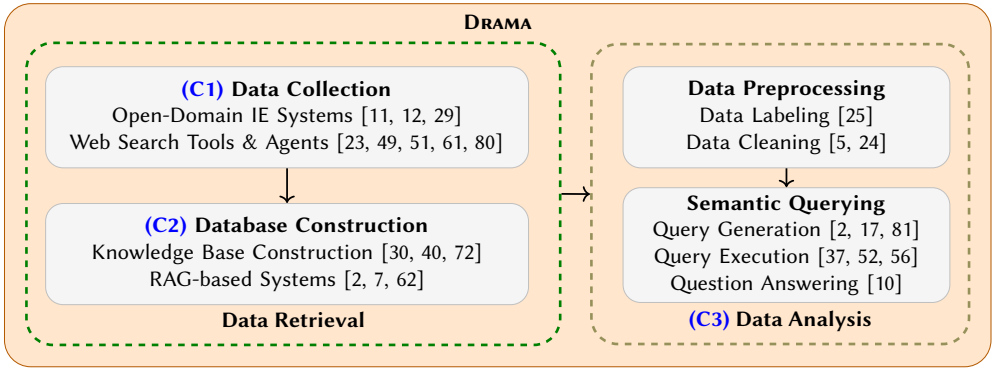

Data science workflows can generally be divided into two phases: data retrieval and data analysis \cite{data-science-life-cycle, Hazzan2023} (Figure \ref{fig:dynamicdb-architecture}).
In real-world scenarios, analysts must navigate data that is both large in volume and rapidly evolving. Successfully completing tasks grounded in such data with high accuracy requires the integration of three key capabilities: \textbf{(C1)} open-domain data collection, \textbf{(C2)} structured data transformation, and \textbf{(C3)} analytic reasoning.

Consider a data analyst manually answering the question: ``What is the national park with the highest visitor spending in 2023 in US?'' \cite{usafacts2024parks} ($Q_0$). To obtain the newly released 2023 data, the analyst needs to first locate and download the relevant and authoritative data from an official government website \cite{npsNature} \textbf{(C1)}.
The collected data is a 68-page PDF report on National Park Visitor Spending Effects \cite{nps2023spending} (i.e., unstructured data). Table 5, spanning pages 27 to 39, contains total visitor spending information for all 433 National Park System (NPS) units, listed in alphabetical order by name. To answer $Q_0$, the analyst further needs to manually extract and digitize this information by, for example, entering the data into a CSV file \textbf{(C2)}.
Finally, the analyst must use analytic reasoning over structured data \textbf{(C3)}. This involves filtering the 63 national parks, which are a subset of NPS units and 
can only be identified by the suffix ``NP'' in the park unit name (e.g., ``Acadia NP'').
The analyst must therefore recognize that this information is embedded in the name itself rather than stored in a separate column and correctly interpret the semantic role of the suffixes.
The analyst must also determine the correct field among seven numeric columns and then identify the maximum value. This process results in a query such as:
\begin{lstlisting}[language=CustomSQL]
SELECT park_unit, total_visitor_spending FROM table
WHERE park_unit LIKE '%NP'
ORDER BY total_visitor_spending DESC LIMIT 1
\end{lstlisting}

Traditional data science workflows are tedious and time-consuming \cite{8440815}, highlighting the critical need for automation. However, existing systems face two key limitations that hinder full automation:

First, from the perspective of end-to-end completeness, 
as we outline in Figure \ref{fig:dynamicdb-architecture}, none of the existing paradigms or systems are capable of simultaneously demonstrating all of \textbf{(C1)}–\textbf{(C3)}.
Specifically, existing open-domain information extraction (IE) systems \cite{10.1145/988672.988687, 10.5555/2283396.2283398, kamp2023openinformationextractionreview} and web search tools \cite{openai2024tools, he2024webvoyagerbuildingendtoendweb, btahir2024open, openai2024researchbot, yang2023autogptonlinedecisionmaking}
actively collect data from the web with no guaranteed accuracy through computations grounded in data.
Existing retrieval-augmented generation (RAG)-based systems \cite{biswal2024text2sqlenoughunifyingai, Chen_2023, tang2024minicheckefficientfactcheckingllms}, even those incorporating dynamic knowledge base construction \cite{liška2022streamingqabenchmarkadaptationnew, wang2023knowledgptenhancinglargelanguage, kasai2024realtimeqawhatsanswer}, assume that all relevant documents have already been collected from the open domains. They do not actively collect up-to-date or task-specific data, and instead operate on a fixed set.
Both open-domain retrieval systems and RAG-based approaches lack support for analytic reasoning over structured data, as their outputs rely on surface-level text generation rather than computations grounded in the underlying data.
Existing data analytic tools \cite{Hu_2024, cao2024spider2vfarmultimodalagents, hong2024datainterpreterllmagent, biswal2024text2sqlenoughunifyingai, zhang2024datacopilotbridgingbillionsdata, dail_sql, patel2025semanticoperatorsdeclarativemodel, liu2024declarativeoptimizingaiworkloads, shankar2025docetlagenticqueryrewriting, eisenschlos-etal-2020-understanding}, on the other hand, support accurate computations over structured data, but assume that the data is readily collected and transformed. 
In fact, except for TAG \cite{biswal2024text2sqlenoughunifyingai}, which unifies RAG \textbf{(C2)} and query generation \textbf{(C3)}, all other existing paradigms and systems focus on only one of the three key capabilities.

Second, although existing paradigms and systems target subsets of \textbf{(C1)}–\textbf{(C3)}, their corresponding capabilities rely on strong and often unrealistic assumptions, and deteriorate when placed within end-to-end settings. Thus, simply aggregating systems with different capabilities is insufficient to automate the full pipeline. 
Specifically, open-domain IE systems and web search tools operate independently of downstream transformation (\textbf{C2}) and analysis (\textbf{C3}), limiting them to simple fact lookups and preventing them from supporting large-scale data collection, thus only partially realizing \textbf{(C1)}.
RAG-based systems presume that data collected from open domains is already clean, structured, and uniform, making them unable to handle data heterogeneity. Their separation from downstream analytic requirements also prevents them from transforming the data to meet analytic needs, such as renaming columns with meaningful labels or aligning schemas. \textbf{(C2)} is thus not fully realized.
The isolation of data analytic tools from upstream data retrieval makes them less flexible to the schema variations, including interpreting contextual cues such as embedded keywords in column values in the previous example, limiting their effectiveness in \textbf{(C3)}.

To address the limitations of existing systems within the scope of tasks that require both (1) large-scale, real-world data collection from open domains and (2) complex analysis and structured reasoning over the collected data to support decision-making,
 we introduce a new end-to-end paradigm \paradigmname, the \textbf{D}ata \textbf{R}etrieval and \textbf{A}nalytical \textbf{MA}nagement pipeline (Section~\ref{sec:paradigmn}). This paradigm answers a user's analytic query $Q$ in natural language with an answer $A$ by integrating data collection, transformation, and analysis.
Specifically, \paradigmname consists of three key stages, as illustrated in Figure~\ref{fig:para}.
First, the data collection stage, \texttt{collect}, collects the necessary raw data $D$ from open domains based on the query $Q$. The collected data can appear in various formats, ranging from Excel files to PDF reports, depending on the nature of the data source.
Second, the data transformation stage, \texttt{transform}, constructs a structured database $T$ based on both $D$ and $Q$. This stage extracts information from raw data and organize it for downstream analytic reasoning.
Third, the data analysis stage, \texttt{analyze}, answers $Q$ over $T$ with $A$ as the final result.
\paradigmname is unified by the following three simple equations and applies broadly to real-world scenarios. 

\begin{align}
\text{Data Collection:} \quad 
& \texttt{collect}(Q) \rightarrow D \\
\text{Data Transformation:} \quad & \texttt{transform}(Q, D) \rightarrow T \\
\text{Data Analysis:} \quad & \texttt{analyze}(Q, T) \rightarrow A
\end{align}

To quantitatively demonstrate system performance on real-world tasks that capture the essential challenges of \paradigmname, we introduce \datasetname in Section~\ref{sec:data}. \datasetname consists of two categories of tasks: (1) claim verification and (2) question answering, each comprising 100 task instances. 
Unlike existing benchmarks, \datasetname is designed to reflect the challenges of performing both large-scale, open-domain data collection and complex analytical reasoning over the collected data, which are not reflected in existing benchmarks. Specifically,
existing open-domain claim verification and question answering benchmarks (1) focus on simple text lookups rather than large-scale, real-world data collection, and (2) evaluate system performance based on textual similarity rather than the analytic reasoning required for decision-making \cite{liu-etal-2023-evaluating,he2024webvoyagerbuildingendtoendweb}. 
Existing data-grounded verification and question answering benchmarks
assume that a static and fully collected data is readily available \cite{li2024can, lei2024spider}, bypassing the challenges of dynamic and large-scale data collection.
As a result, no existing benchmarks adequately capture the scope or application scenarios that \paradigmname is designed to address.
To fill this gap, \datasetname provides tasks drawn from content dated on or after January 1st, 2024, which is later than the knowledge cutoffs of state-of-the-art large language models (LLMs). This ensures that completing them requires collecting data from open domains.
The claim verification tasks consist of real-world false claims posted on social media that have been publicly corrected using verifiable, publicly available data. The question answering tasks involve analytic factual questions addressing socially and economically impactful topics, with deterministic answers grounded in public data sources. This ensures that completing the tasks requires constructing structured databases and performing precise analysis.

We propose \sysname in Section \ref{sec:method}, a multi-agent system built on top of \paradigmname, comprising a data retriever and a data analyzer.
The data retriever coordinates the execution of three sub-agents: web browser, data transformer, and web augmenter. The web browser collects raw data $D$ through fine-grained webpage access. The data transformer then constructs a structured database $T$ by extracting and organizing relevant information from $D$. If $T$ does not include sufficient information, the web augmenter is triggered to collect $D$ from a broader range of webpages in large quantities.
The data analyzer defines and executes a function $f(Q, T) \rightarrow A$, where $T$ is a structured table and $A$ is the final answer to the user’s query $Q$.

We evaluate \sysname together with five baseline agents 
on \datasetname. 
\sysname achieves 86.5\% task accuracy at a cost of \$0.05, outperforming all baselines with the highest accuracy and lowest API cost.  
The accuracy of \sysname is $6.9\times$ that of the recently released OpenAI Research Agent \cite{openai2024researchbot}, while incurring less than 1/6 of the cost.
\sysname also demonstrates stable performance across different task categories and types.
We report the details of these evaluations in Section \ref{sec:exp}.

The contributions of this paper can be summarized as follows:

\begin{itemize} 
\item We propose an end-to-end paradigm, \paradigmname, that unifies data collection, transformation, and analysis.
\item We collect a benchmark, \datasetname, consisting of real-world tasks that are representative of \paradigmname.
\item We develop a multi-agent system, \sysname, that effectively operationalizes \paradigmname. 
\item We quantitatively evaluate the performance of \sysname and state-of-the-arts agentic systems on \datasetname. 
\end{itemize}
\section{\paradigmname: A Paradigm that Unifies Data Collection, Transformation, and Analysis}\label{sec:paradigmn}

In this section, we introduce the \paradigmname paradigm, a pipeline that takes as input a user query $Q$ and returns an answer $A$ by retrieving and analyzing data from open domains.
We outline three major stages of \paradigmname: data collection (Section~\ref{subsec:para_data_retrieval}), data transformation (Section~\ref{subsec:para_data_transformation}), and data analysis (Section~\ref{subsec:para_data_analysis}).
For each stage, we illustrate the expected system performance, outline the necessary capabilities, and highlight the limitations of existing systems to demonstrate why \paradigmname cannot be achieved through a trivial integration of existing systems and engineering effort.
While we present \paradigmname as executing a single iteration of these stages, it can be naturally extended to a multi-hop setting, including partially multi-hop workflows where, for example, the system iterates between data collection and data transformation before proceeding to analysis (e.g., $(1\rightarrow2)_n\rightarrow3$ in Section \ref{sec:method}), for more complex tasks. 

We use two motivating examples of real-world tasks that require retrieving up-to-date data and demand accuracy: one example is answering the question: ``Which (USA) state has the highest rate of homelessness in 2024?''\cite{usafactsHomelessness2024} ($Q_1$). Another example arises from the proliferation of claims on social media, where individuals must quickly distinguish truth from misinformation, such as verifying: ``Change in the number of births from 2022 to 2023 for Ireland: -10.3\%''\cite{xPost1821510949771059296} ($Q_2$).

\subsection{Data Collection}\label{subsec:para_data_retrieval}
The \texttt{collect} function takes a user query $Q$ in natural language as input and searches for relevant raw data $D$ from open domains to answer $Q$.
We present two examples in Figure~\ref{fig:para}, where to answer $Q_1$ and verify $Q_2$, the \texttt{collect} function browses open domains, identifies websites containing the most relevant and authoritative data, retrieves the raw data, and caches it as an intermediate result.
Note that $D$ can appear in various formats depending on the data source. In Figure~\ref{fig:para}, we see that for $Q_1$, the raw data $D_1$ is an Excel sheet \cite{hudAHAR2024}, while for $Q_2$, $D_2$ is a report in PDF format \cite{csoVitalStats2023}.

Existing automated data science tools lack the capability to perform open-domain data search and retrieval. This functionality cannot be trivially achieved by combining information extraction or web search tools, as these systems are primarily designed for text-based lookups with limited context, rather than retrieving large-scale data from diverse sources and formats, as required by the \texttt{collect} function.

\subsection{Data Transformation}\label{subsec:para_data_transformation}
The retrieved data $D$ can appear in multiple formats, ranging from unstructured data like PDFs to structured data like CSVs. In this stage, the \texttt{transform} function extracts the required information from $D$ and organize it as a structured database $T$. Specifically, \paradigmname adopts a single-table representation in the data transformation step to enable more reliable execution of complex queries. Prior work has shown that directly querying over multi-table databases is already challenging and error-prone in closed-domain settings \cite{lei2025spider20evaluatinglanguage}, and this difficulty is amplified in open-domain scenarios where data is less structured and consistent. Importantly, the process of integrating multiple tables often reveals the need for additional transformations that are not obvious when viewing each table in isolation, such as pivoting, transposing, unpivoting, or invoking user-defined functions. Such transformations are cumbersome and unnatural to express directly within downstream SQL query generation, and are more effectively handled in a dedicated transformation step prior to analysis. 

As we show in Figure~\ref{fig:para}, $D_1$ is an Excel sheet containing 18 tabs, each with 1,306 columns. The \texttt{transform} function selects the relevant information, specifically the tab for the year 2024 and two columns (\textit{State} and \textit{Overall Homeless}), and produces $T_1$, a structured table containing these columns. Similarly, $D_2$ is a PDF report containing statistics on births, deaths, and natural increases from 2013 to 2023. The \texttt{transform} function extracts the birth data for 2022 and 2023 from a diagram and organizes it as $T_2$.

As an intermediate component between open-domain data collection and data analysis, the \texttt{transform} function must process data in diverse formats, parse them into structured tables, understand their relationships, and present them in forms that are interpretable and suitable for downstream analysis. Existing systems lack the capability to automatically handle such heterogeneous data formats in a unified and generalizable manner.

\subsection{Data Analysis}\label{subsec:para_data_analysis}
\begin{figure}[htbp!]
    \graphicspath{{figures/}}
    \centering
    \includegraphics[width=0.92\columnwidth]{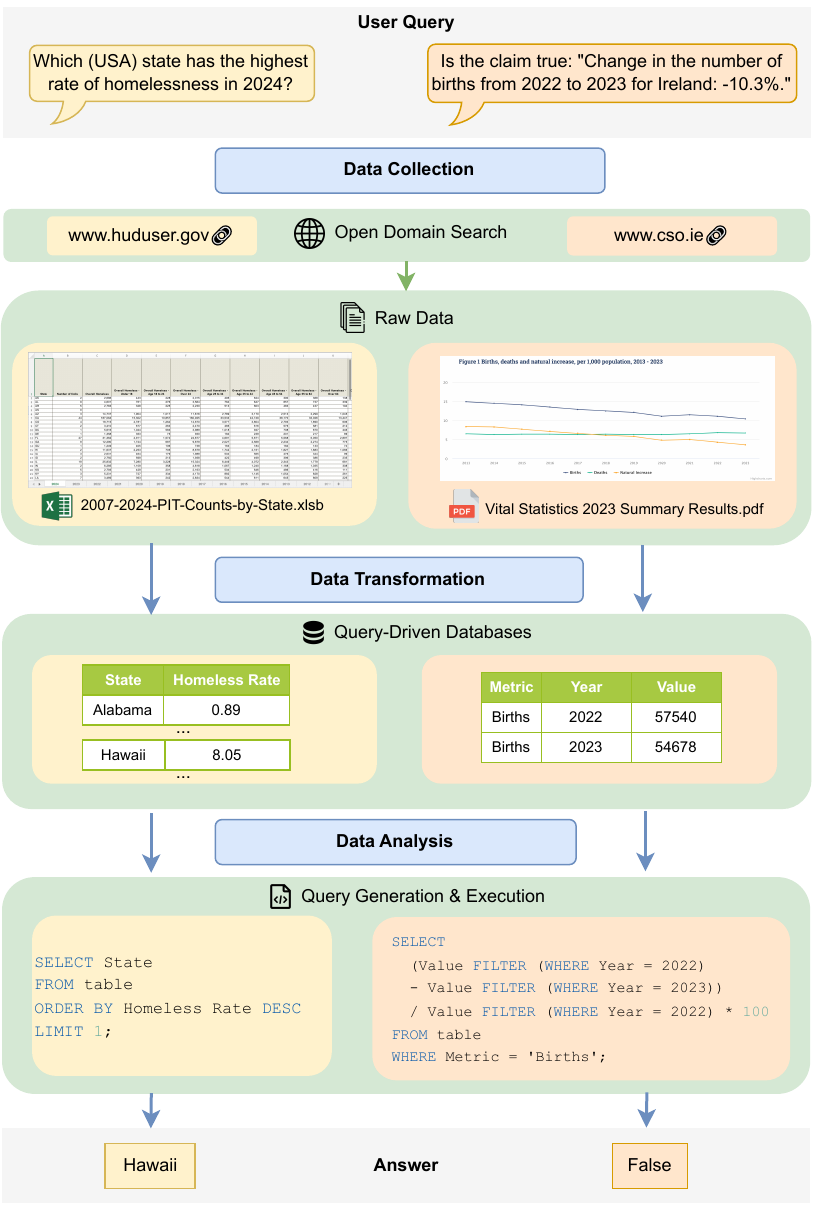}
    \caption{Overview of the \paradigmname paradigm. Here we present two examples: \textit{(left)} user query as a question ($Q_1$), and \textit{(right)} user query as a claim to be verified ($Q_2$).}
    \label{fig:para}
\end{figure}

This stage abstracts the task of answering a natural language query $Q$ over a structured table $T$ as the \texttt{analyze} function. In Figure~\ref{fig:para}, we demonstrate one approach: translating $Q$ into SQL operations and executing the corresponding code. However, this stage is flexible and can be adapted to any other state-of-the-art semantic query parsing and execution systems, including translating $Q$ into any other programming languages such as into Python code with pandas functions or using table question answering frameworks.

While the \texttt{analyze} function can be realized by a wide range of natural language query systems over structured tables, it also requires adaptation to handle the diverse and complex formats and content structures produced during earlier stages of the pipeline.

\section{\datasetname: A Collection of Real-world \paradigmname Applications} \label{sec:data}
\begin{figure}[htbp!]
    \graphicspath{{figures/}}
    \centering
    \includegraphics[width=0.95\columnwidth]{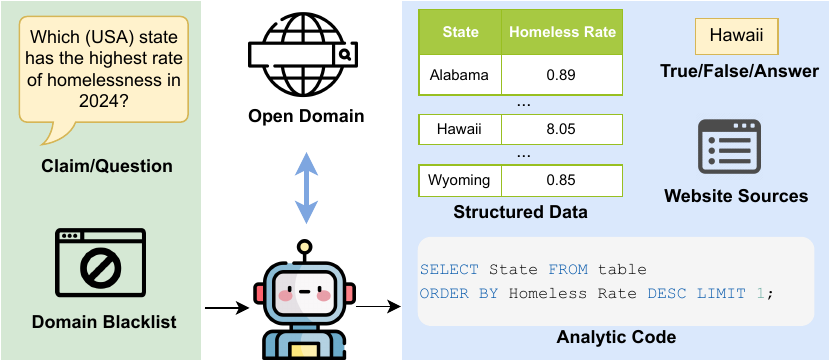}
    \caption{Overview of each \datasetname task. Given a user query, the agent is tasked with collecting, structuring, and analyzing data from open domains to generate an answer.}
    \label{fig:task}
\end{figure}

To quantitatively assess system performance on real-world tasks that faithfully reflect the core features of \paradigmname, we construct \datasetname, a benchmark comprising two task categories: (1) claim verification and (2) question answering, each with 100 task instances.
We describe the task collection criteria and process in Section~\ref{subsec:data-collection} and formally define the tasks in Section~\ref{subsec:task-desc}.

\begin{table}[h]
\centering
\small
\caption{Examples of \datasetname data}
\begin{tabular}{p{0.1\textwidth} p{0.77\textwidth} p{0.06\textwidth}}
\toprule
\textbf{Task} & \textbf{Claim/Question Example} & \textbf{Label} \\
\midrule
\multirow{2}{*}{\shortstack[l]{Claim\\Verification}}
& Change in the number of births from 2022 to 2023 for Ireland: -10.3\% \cite{xPost1821510949771059296} & False \\
& Change in the number of births from 2022 to 2023 for Ireland: -5\% \cite{xCommunityNote1821558069928620304} & True \\
\midrule
\multirow{2}{*}{\shortstack[l]{Question\\Answering}}
& Which state has the highest rate of homelessness in 2024? \cite{usafactsHomelessness2024} & Hawaii \\
& What is the average number of wildfires in the U.S. annually from 1983 to 2023? \cite{usafactsWildfiresDamage2024} & 70{,}000 \\
\bottomrule
\end{tabular}
\label{tab:task-exp}
\end{table}

\subsection{Task Collection} \label{subsec:data-collection}

To ensure that \datasetname reflects the core characteristics of \paradigmname-style tasks and enables generalizable evaluation of agent performance in realistic, paradigm-aligned scenarios, we apply the following universal criteria to both task categories: (1) the claim/question should be real-world and important (i.e., publicly posted on influential social media platforms and significant enough to warrant verification or answering); (2) it should be analytic in nature (i.e., based on data analysis results); (3) it should be non-vague (i.e., clearly and deterministically mappable to the execution result of a SQL script); (4) it should be based on publicly available and verified data, with explicit reference to the data. 

We further introduce task-wise data collection criteria as follows:

\minihead{Claim Verification}
In this task, the agent is required to retrieve data from open domains to verify whether a claim is true or false.
To ensure a fair and balanced comparison, we require that true and false claims exhibit:
(1) a comparable number of instances, and
(2) comparable difficulty levels, both in data retrieval and data analytics.
To do this, we source claims from PolitiFact \cite{usafacts} and Twitter/X Community Notes \cite{communitynotes_data}.
As a task-specific criterion, we require that the falsity in claims must originate from incorrect data analysis.

For PolitiFact, we extract all 18 fact-check posts with ratings equal to or lower than Mostly False, published between 01/01/2024 and 02/28/2025, that satisfy the universal and task-specific criteria.
From each post, we collect the original claim (as the false claim) and the accompanying explanation (as the true claim).

For Twitter/X Community Notes, we download all notes available as of 02/28/2025 and apply the following filters:
(1) the original Twitter/X post is classified as \texttt{MISINFORMED\allowbreak\_OR\allowbreak\_POTENTIALLY\allowbreak\_MISLEADING};
(2) the post contains a \texttt{misleadingFactualError};
(3) the note refers to \texttt{trustworthy \allowbreak Sources};
(4) the \texttt{currentStatus} of the community note is \texttt{CURRENTLY\_RATED\_HELPFUL};
(5) the post contains no media contents (e.g., images or videos);
(6) the note includes the keyword \texttt{data};
(7) both the post and the note are in English; and
(8) to ensure data stability due to the fast pace and loosened constraints of community notes, we restrict collection to notes posted between 01/01/2024 and 01/01/2025, ensuring a minimum 3-month gap from the time of collection.
This filtering yields 545 pre-filtered notes.
From these, we manually extract 32 that satisfy all universal and task-specific criteria.
We treat the Twitter/X post as the false claim and the community note as the true claim.

Our pairing strategy in collecting claim verification tasks ensures a balanced number and comparable difficulty, as both claims are based on the same dataset and share the same underlying SQL logic.
We provide example true and false claims in Table~\ref{tab:task-exp}.

\minihead{Question Answering}
In this task, the agent must retrieve data from open domains to answer a given question.
To ensure reliable evaluation, we impose the following task-specific criterion: the answer must be representable as a single string or numeric value resulting from data analysis, such as the output of a SQL query or an equivalent computation.

We collect all 100 valid statements from the Economy and Education sections of USAFacts \cite{usafacts2025}, published between 01/01/2024 and 02/28/2025, that satisfy the universal and task-specific criteria.
We use GPT-4o \cite{openai2024gpt4o} to rephrase each statement into a question and then manually annotate the ground-truth answers based on the original USAFacts article.
Table~\ref{tab:task-exp} includes examples with both text and numeric answers. 
There are 31 task instances with text answers and 69 with numeric answers as ground truths.

\vspace{.5em}
For each task instance for both verification and QA tasks, we collect the ground-truth answer, manually transform the original source data into a structured table, and annotate SQL scripts that reproduce the result. This enables fine-grained evaluation of both data retrieval and analysis performance. 
Specifically, all 200 task instances were annotated by a team of 5 researchers, each with substantial SQL experience. The annotation process consisted of two stages:

\begin{enumerate}
\item Each annotator independently transformed the raw data into structured form and generated the corresponding SQL code in response to the natural language question or claim. At this stage, each data-code pair was executed to ensure it produced the expected ground-truth answer (i.e., True/False for verification tasks or the exact value for QA).
\item The remaining four authors then independently cross-verified the data transformation and SQL code. Any discrepancies were resolved collaboratively until full agreement was reached on the final version.
\end{enumerate}

To prevent data leakage, we compile the original sources into a blacklist, which is provided to the agents at execution time. Agents are explicitly restricted from accessing any content in the blacklist during task execution. We verify that, although not explicitly constrained, for all 200 out of 200 tasks, all data required to answer each query, while potentially spanning multiple files or tables, comes from a single domain by the nature of the tasks.

\subsection{Task Definitions} \label{subsec:task-desc}
As we illustrate in Figure~\ref{fig:task}, we formally define the \datasetname tasks as follows: given (1) a claim to be verified or a question to be answered, and (2) a blacklist of domains that must not be used as data sources, the system is required to retrieve relevant data from open domains (e.g., the web) and analyze it to output: (1) the validity of the claim or the answer to the question, (2) the retrieved data, transformed into a structured table, (3) the analytic code used to derive the result, and (4) traces (i.e., the websites accessed during the data collection process).
\section{\sysname: Realizing \paradigmname Through a Multi-Agent System}\label{sec:method}
\begin{figure}[htbp!]
    \graphicspath{{figures/}}
    \centering
    \includegraphics[width=\columnwidth]{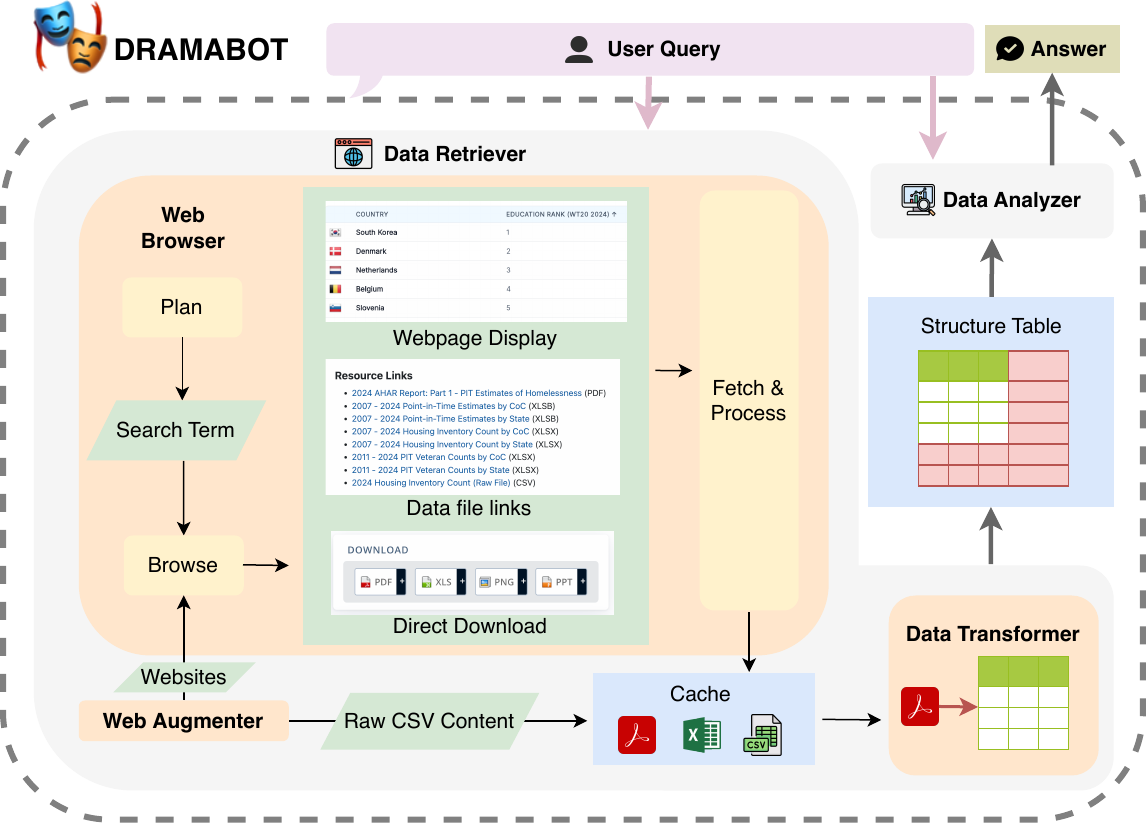}
    \caption{Overview of \sysname.}
    \label{fig:sys}
\end{figure}
In this section, we develop \sysname following \paradigmname. \sysname is a multi-agent system that is capable of executing the end-to-end data science pipeline consisting of a data retriever (Section \ref{subsec:data_retriever}) that executes stages 1 and 2, data collection and data transformation, and a data analyzer (Section \ref{subsec:data_analyzer}) that executes stage 3, data analysis. 

Following the \datasetname task interface, \sysname takes as input a user query $Q$ and a blacklist of domains $B$ that should not be accessed, and returns: the answer $A$ to the query, the structured table $T$ containing information to answer the query, the set of sources $S$ used to obtain the data, and the code program $P$ executed on $T$ to generate $A$. 
\sysname ensures that no source $s \in S$ is drawn from $\mathcal{B}$.

\subsection{Data Retriever}\label{subsec:data_retriever}
The goal of the data retriever is to: \textbf{(G1)} collect data $D$ sufficient to answer $Q$, \textbf{(G2)} ensure the data comes from reliable sources $S$ that are not associated with $B$, and \textbf{(G3)} transform $D$ as a structured table $T$ tailored to $Q$, stored in \sysname environment shared with data analyzer.

The data retriever coordinates the execution of three sub-agents: web browser (Section \ref{subsec:web_browser}), data transformer (Section \ref{subsec:data_manager}), and web augmenter (Section \ref{subsec:web_augmenter}). We provide an overview of the workflow in Figure \ref{fig:sys}.
The data collection stage begins with an empty set of sources $S = \emptyset$, and progressively builds $S$ as the data retriever coordinates the web browser and the web augmenter as two complementary modules.
The web browser performs detailed, step-by-step exploration to collect data from a small and precise set of sources, aiming to identify the most reliable domain. If it fails to collect sufficient information from authoritative sites, the web augmenter retrieves a broader set of candidate data from less curated sources in large quantity. This data is then inspected based on both source reliability and content relevance to ensure accuracy.
Through this coordination, the data retriever ultimately selects a single, most trustworthy domain and retrieves all required data from that source.
The data transformation stage is performed by the data transformer, where the data collected by the web browser and the web augmenter is aggregated and organized. 
The key distinction of \sysname from existing systems lies in its ability to collect large-scale data across a wide range of formats using the web browser, enabled by its uniquely designed interfaces with open domains and coordination with the web augmenter. Additionally, the data transformer plays a critical role in converting this heterogeneous data into a single, descriptive structured table suitable for downstream analysis.

We now describe the structure, communication, and invocation conditions of each agent in detail. 

\subsubsection{Web Browser} \label{subsec:web_browser}
The web browser takes $Q$ and $B$ as input and collects raw data $D$, stored in cache as intermediate results. A starting website, $W$, can also be provided as input; by default, $W = \emptyset$, meaning the web browser starts on the \texttt{google.com} page and generates an initial search term to guide the search path for data retrieval. 

The browsing process is adapted from WebVoyager \cite{he2024webvoyagerbuildingendtoendweb}, a general-purpose web agent. The web browser uses Selenium \cite{selenium2025}, and at each step, the agent receives a screenshot of the webpage as an observation of the environment, along with feedback from previous steps. For every iteration, the web browser selects an action from its action space and executes it in the web environment. Before selecting an action in each round, the URL of the driver task for that round is added to $S$.

We further make the following modifications to its original action space, which includes \texttt{Click}, \texttt{Type}, \texttt{Scroll}, \texttt{Wait}, \texttt{GoBack}, \texttt{Google}, and \texttt{ANSWER}:
\begin{enumerate}
    \item Modify the \texttt{Click} action: this action only clicks elements that are not associated with any domain in $B$.
    \item Remove the \texttt{Answer} action.
    \item Add \texttt{GetData} action: this action completes the browsing process by returning the specific data contents in CSV format and should be performed when the agent finds the necessary data to answer $Q$. 
    \item Add \texttt{CheckLink} action: this action checks the hyperlink of a web element.
    \item Add \texttt{GetLink} action: this action completes the browsing process by returning the hyperlink to the data file. 
    \item Add \texttt{Download} action: this action monitors the data file download process and completes the browsing process if the data file is successfully downloaded. 
\end{enumerate}

Adjustment (1) supports \textbf{(G2)} by preventing the web browser from accessing unreliable sources. Adjustments (2)–(6) support \sysname's ability to meet \textbf{(G1)} by enabling large-volume data exchange between open domains and the local environment, overcoming the limitations of existing web search tools. Specifically, we replace the simple textual response (Adjustment (2)) with the three data collection mechanisms (Adjustments (4)–(6)) to accommodate the following variety in data sources:

First, data can be directly collected from webpages (Adjustment (3)). For example, as we show in Figure~\ref{fig:sys}, to verify the claim ``The USA ranks \#1 in education rankings in Education Rankings by Country 2025,''\cite{xPost1769443227902341380} the relevant rank data is directly displayed on the webpage \cite{wprEducationRankings}.
Second, data can be collected via direct hyperlinks exposed in the webpage HTML (Adjustments (4) and (5), which operate as a pair). For example, as we show in Figure~\ref{fig:sys}, to answer $Q_1$, the hyperlinks to the relevant data files are visible and embedded directly in the page \cite{hudAHAR2024}.
Third, data can be collected from files that are served through backend JavaScript pipelines, which makes them difficult to extract directly from the HTML content. For example, as we show in Figure~\ref{fig:sys}, verifying the claim ``India average cigarettes inhaled daily: 8''\cite{xPost1858383505501098277} requires clicking a download button \cite{statistaIndiaPollution} and waiting for the file to finish downloading before the relevant data becomes accessible.

Corresponding to the action space, we provide a list of updated action guidelines to ensure accurate and consistent behavior of the web browser. Below, we present four critical instructions:

\begin{enumerate}
\item \texttt{GetData} should 
browse the webpage incrementally, view by view, and gradually accumulate raw data into a structured table with clear and straightforward column names until sufficient information is collected. This strategy is supported by the promising performance of existing multimodal large language models (MLLMs) on data extraction tasks \cite{openai2024gpt4o-2}, which enables reliable extraction from each individual view (i.e., webpage screenshots).

\item \texttt{CheckLink} must be executed before performing \texttt{GetLink} on the same element. This ensures that all returned hyperlinks are verified as pointing to valid data files.

\item The web browser should always scroll up and down to view the entire webpage before deciding to navigate to other pages. 
This ensures that the web browser thoroughly explores each page before proceeding.

\item The web browser should prioritize retrieving data from authoritative sources, such as official government websites, and avoid low-quality or unofficial domains. This helps prevent misinformation from unverified sources.
\end{enumerate}

The browsing process returns collection targets $R$ for raw data $D$, which appear as (1) raw contents (via \texttt{GetData}), (3) hyperlinks to data files (via \texttt{GetLink}), and (3) original data files (via \texttt{Download}). The web browser further processes $R$ to store $D$ as cache in agent environment. Specifically, the web browser issues the following shell commands via conventional software pipelines:
\begin{enumerate}
  \item \textbf{Raw content:} Dump the raw contents $R$ as a CSV file in the agent environment:
  
\begin{lstlisting}[style=cmdstyle]
echo $R > /agent_env/cached_data.csv
\end{lstlisting}

  \item \textbf{Hyperlink:} Fetch the file from the hyperlink $R$ using \texttt{curl}, and unzip if necessary:

\begin{lstlisting}[style=cmdstyle]
curl -OJ $R --output-dir /agent_env/
unzip /agent_env/*.zip
\end{lstlisting}
  
  \item \textbf{Original file:} Move the downloaded file $R$ from the download directory into the agent environment:

\begin{lstlisting}[style=cmdstyle]
mv /downloads/$R /agent_env/
\end{lstlisting}

\end{enumerate}

With $D$ collected, the web browser concludes its task round and returns control to the data retriever, which then invokes the data transformer (Section \ref{subsec:data_manager}).

\subsubsection{Data Transformer} \label{subsec:data_manager}
The data transformer performs stage 2 of \paradigmname, where the raw data $D$ is transformed into a structured table $T$ based on $Q$ for subsequent analysis to support \textbf{(G3)}.

Before introducing the detailed operations carried out by the data transformer, we define its core function: \texttt{aggregate\_tables($T_1$, $T_2$, $Q$, \dots) $\rightarrow T_{\text{new}}$}. 
Unlike traditional SQL join methods that follow fixed syntax and join logic, this function supports a variety of merge strategies that adapt dynamically to $Q$ and other given contexts. We summarize three basic patterns of table aggregations used when constructing $T$. We use $C$ to denote columns and $R$ to denote rows. Specifically,
\begin{itemize}
    \item \textbf{Column Aggregation}: Combines attributes from multiple tables along shared rows based on a common key.  
    For example, $T_1$ contains homelessness rates by state, $T_2$ contains housing inventory counts by state, and $Q$ asks for the correlation across the factors in the two tables \cite{usafactsHomelessness2024}. In this case, joining $T_1$ and $T_2$ is equivalent to performing a SQL join on the shared \texttt{State} column:
    \begin{lstlisting}[language=CustomSQL1]
SELECT * FROM T1
JOIN T2 ON T1.State = T2.State
    \end{lstlisting}

    \item \textbf{Row Aggregation}: Merges rows from two tables with the same schema to form a single dataset spanning a larger range.  
    For example, $T_1$ contains the birth rates in Ireland from 2003–2013, $T_2$ contains the same data from 2013–2023, and $Q$ asks for trends across these two decades \cite{xPost1821510949771059296}. In this case, joining $T_1$ and $T_2$ involves stacking $T_2$'s rows below $T_1$'s, corresponding exactly to \texttt{UNION} or \texttt{UNION ALL} in SQL:
    \begin{lstlisting}[language=CustomSQL1]
SELECT Year, BirthRate FROM T1
UNION ALL
SELECT Year, BirthRate FROM T2
    \end{lstlisting}

    \item \textbf{Mixed Aggregation}: Combines row-wise merging with the addition of distinguishing metadata to track source-specific attributes.  
    For example, $T_1$ contains homelessness rates by state for 2023, $T_2$ contains the same for 2024, and $Q$ asks to compare the rates across the two years \cite{usafactsHomelessness2024}. To support this, a new column \texttt{Year} is added, with 2023 assigned to rows from $T_1$ and 2024 to rows from $T_2$. All rows are concatenated and preserved in $T_{\text{new}}$:
    \begin{lstlisting}[language=CustomSQL1]
SELECT state, rate, 2023 AS year FROM T1
UNION
SELECT state, rate, 2024 AS year FROM T2
    \end{lstlisting}
\end{itemize}

In addition to following these basic patterns, the \texttt{aggregate\_tables} function is also instructed to apply transformations that facilitate downstream analysis, such as renaming columns with contextually meaningful names, based on the specified context, including $Q$.

We now illustrate the workflow of the data transformer along with the example from Section~\ref{sec:intro}, with the National Park Visitor Spending Effects report downloaded locally as a PDF, as part of $D$.
The data transformer maintains a list $L$ to track the files it has already inspected. To assist in decision-making, the contents of any file whose name contains \texttt{readme} (case-insensitive) are first parsed and incorporated into the context. Once parsed, these readme files are added to the checked file list $L$. Starting with $T = \emptyset$, the transformer performs the following operations in each iteration:

\minihead{\texttt{check\_adequate\_info($Q$, $T$) $\rightarrow$ \{(True, $C$), (False, $M$)\}}} This operation checks the adequacy of existing data through a code-driven inspection process. The data transformer loads the current local data storage $T$ and determines whether it contains sufficient information to answer the query $Q$. To do this, it attempts to generate an executable code snippet based on $T$ to answer $Q$. If a deterministic code snippet can be generated, this indicates that $T$ is adequate. The data transformer then completes the current task round and returns $C$ as a reference for coordination with other agents in \sysname. If not, it outputs a list of missing information $M$ to guide future decisions on file selection and information extraction, i.e., the proceeding operations.
In the first iteration of our example, this operation returns \texttt{False} because $T$ is empty, identifying the park names and total visitor spending as the missing information $M$.

\minihead{\texttt{file\_selection($Q$, $M$, $D$, $L$) $\rightarrow$ $F$}} 
This operation selects a data file $F$ from the set of unprocessed files $D \setminus L$ that is most likely to contain the missing information $M$.
The data transformer is presented with a list of file names. 
In our example, the National Park Visitor Spending Effects report file is selected because its name is the most relevant and likely to contain the missing information.

\minihead{\texttt{extract\_data($Q$, $M$, $F$, $L$) $\rightarrow$ $T'$}} This operation extracts the information necessary to answer $Q$, particularly the missing elements in $M$, from a data file $F$, and formulates it as a tentative structured table $T'$. 
We categorize the following two extraction approaches based on the data type of $F$:
\begin{itemize}
    \item \textbf{Structured data} (e.g., \texttt{.csv}, \texttt{.tsv}) and \textbf{Semi-structured data} (e.g., \texttt{.xls}, \texttt{.xlsx}): 
    The structured units of $F$ (e.g., the entire CSV file, or each sheet of an Excel file) are first extracted as ${T_1, T_2, \dots, T_n}$. These tables are then iteratively joined using \texttt{aggregate\_tables} with both $Q$ and $M$ passed in contexts: at each step $i$, the intermediate result $T'_i$ is updated as $T'_i = \texttt{aggregate\_tables}(T'_{i-1}, T_i, Q, M)$, starting with $T'_0 = \emptyset$ for $T_1$, including the sole structured data file, to apply transformations. 
    \item \textbf{Unstructured data} (e.g., \texttt{.pdf}): 
    Given the robust performance of MLLMs in structured data extraction from various document types \cite{biswas2024robustnessstructureddataextraction}, we use MLLMs (GPT-4o \cite{openai2024gpt4o}) for data extraction. However, since MLLM performance tends to degrade with increased input volume \cite{liu-etal-2024-lost}, our core strategy is to extract information from small, manageable portions of data incrementally and then aggregate the results. 
    Specifically, the minimal processing unit of $F$ (e.g., a page from a PDF) is sequentially passed to the MLLM. Starting with an empty dataframe $T' = \emptyset$, the MLLM is instructed to iteratively expand $T'$ with the necessary information to answer $Q$ and addresses the information shortage included in $M$, stopping once all required information is captured or the entire file has been processed. To overcome the context limitations of MLLMs, $T'$ is included in each API call as context. To overcome the form-generation limitations of MLLMs, the model is instructed to return $T'$ containing only the minimal information necessary and relevant to answering $Q$.
\end{itemize}

At the end of this operation, before returning $T'$, $F$ is added to $L$.
In our example, since the report is a PDF file, the MLLM is invoked on each page. As the first 26 pages do not contain relevant information, $T'$ remains empty. Starting from page 27, where the table first appears, $T'$ is incrementally enriched with the table contents (i.e., the rows containing park and visitor spending information), continuing until the table concludes at page 39, the operation would return and add the report to $L$.

\minihead{\texttt{update\_database($Q$, $M$, $T$, $T'$) $\rightarrow$ $T$}} In this operation, the extracted tentative structured table $T'$ is joined with $T$ to enrich the available information. Specifically, $T$ is updated as $T = \texttt{aggregate\_tables}(T, T', Q, M)$.
In our example, $T'$ extracted from the report is aggregated with an empty table, where its column names are reformatted to include necessary dashes or underscores to facilitate downstream analysis.

Once all files have been checked in all iterations, the data transformer concludes its task round and returns control to the data retriever. The data retriever then verifies whether $T$ is nonempty and valid, i.e., whether it contains sufficient information to answer $Q$. If this condition is met, 
the data retriever concludes its task and hands control over to the data analyzer (Section~\ref{subsec:data_analyzer}). If not, the data retriever invokes the web augmenter (Section~\ref{subsec:web_augmenter}), triggering additional iterations of stages 1 and 2 of the \paradigmname paradigm. 
In our example, when \texttt{check\_adequate\_info} is called in the second iteration, it returns \texttt{True} together with the SQL code snippet we show in Section~\ref{sec:intro}, concluding the process.

\subsubsection{Web Augmenter} \label{subsec:web_augmenter}
The main goal of the web augmenter is to complement the fine-grained searches of the web browser, where large crowds of sources are fetched within a short timeframe using parallel workloads to support \textbf{(G1)}. Specifically, the web augmenter uses the OpenAI search tool \cite{openai2024tools}, and given $Q$ with access restrictions to $B$, it is tasked with searching the websites to produce data $D'$ containing specific data contents in CSV format using clear and straightforward column names that are necessary to answer $Q$. 
The sources $S'$ accessed by the web augmenter are included in the \texttt{annotations} field of the response.

Although explicitly prompted not to access domains in $B$, when using an integrated tool that doesn't strictly follow the prompts, the web augmenter may still access them. Thus, to support \textbf{(G2)}, when the web augmenter concludes its task round and returns control to the data retriever along with its collected outputs, the data retriever must decide whether to trust the data collected from these large-crowd sources based on their $S'$, or initiate another round of web browsing using the reliable sources \( S'' = S' \setminus B \). The data retriever processes the outputs of the web augmenter as follows:

If $s \notin B, \forall s \in S'$, 
the sources $S$ are augmented with $S = S \cup S'$ and the data in cache is augmented with $D = D \cup D'$. The data transformer is then called to process $D'$, and once it returns $T$, the data retriever concludes its task.

If \( \exists s \in S' \, \text{s.t.} \, s \in B \), the data retriever performs the following steps:

\begin{enumerate}
    \item Excludes all the sources associated with the blacklisted domains by constructing \( S'' = S' \setminus B \).
    \item The data retriever calls the \texttt{rank\_website} function to rank \( S'' \) from the most relevant and reliable sources to answer \( Q \) to the least, based on \( Q \) and the response of the web augmenter, including the result, data, and code, with inline annotations where each \( s \in S'' \) is used to generate the response. \texttt{rank\_website} ranks the reliability and relevance of websites according to their contributions to the response to the query generated by the Web Augmenter.
    \item The web browser is called iteratively with the sources \( s \in S'' \) passed as the starting website \( W \), following the order of the website rankings. For each iteration \( i \):
    \begin{itemize}
        \item \( D \) is expanded with the newly collected data \( D_i \) as \( D = D \cup D_i \), and \( S \) is expanded with the new sources accessed as \( S = S \cup S_i \).
        \item The data transformer is called to process $D_i$. Once the data transformer outputs valid storage \( T \), the data retriever concludes its task, and \sysname calls the data analyzer (Section \ref{subsec:data_analyzer}). 
    \end{itemize}
\end{enumerate}

\subsection{Data Analyzer}\label{subsec:data_analyzer}

With a structured table $T$ stored in local storage, the data analyzer further answers $Q$ over $T$.  When \( T = \emptyset \), the analyzer falls back to return \texttt{False} for verification tasks, indicating that no data are available to support the claim and defaulting to an educated guess, which in many real-world verification scenarios corresponds to labeling unsupported claims as false. When \( T \neq \emptyset \), we implement an NL2SQL system by prompting \texttt{gpt-4o} \cite{openai2024gpt4o} as the query generator. 
Building on the strong performance of existing LLMs on coding tasks \cite{chen2021evaluatinglargelanguagemodels}, our key innovation lies in making the model aware of diverse and potentially noisy data retrieved from open-domain sources. The query generator takes the table head as input to observe and adapt to unconventional or noisy data contents through few-shot learning \cite{brown2020languagemodelsfewshotlearners}.
The generated code is executed over $T$ to produce the answer $A$ generated by \sysname in response to $Q$.

\section{Experiments}\label{sec:exp}
We introduce the experimental setup in Section~\ref{subsec:exp_setup}. We report the overall system-level performance in Section~\ref{subsec:overall_evaluation}, and examine their generalizability in Section~\ref{subsec:generalizability}. We analyze the stage-wise performance of all the agents in Section~\ref{subsec:component_studies} and evaluate the impact of \sysname's subagents, including ablation studies, in Section~\ref{subsec:ablation_studies}.

\subsection{Experiment Setup} \label{subsec:exp_setup}
We evaluate the performance of \sysname and five baseline agents on \datasetname.
\subsubsection{Baselines}\label{subsec:baselines}
We select the following five baseline agents that have demonstrated strong performance across a variety of real-world tasks: 

\minihead{Deep Research \cite{btahir2024open}} Deep Research is an agent that integrates planning, searching, reasoning, and reporting, and has shown strong performance on research tasks involving the collection and analysis of complex document corpora. 
We select Deep Research to examine how its capabilities generalize to tasks requiring the retrieval and analysis of complex, large-scale data. 
We use the open-sourced version of the Gemini Deep Research reasoning framework \cite{google_deep_research_2024}.

\minihead{AutoGPT \cite{AutoGPT}} AutoGPT is a multi-functional agent designed to handle a wide range of tasks, featuring capabilities such as web search, long-term planning, tool selection and usage, and self-reflection. 
We include AutoGPT to evaluate how a general-purpose agent performs on \datasetname in comparison to \sysname.

\minihead{WebVoyager \cite{he2024webvoyagerbuildingendtoendweb}} WebVoyager is an MLLM-powered web agent capable of executing user instructions end-to-end by interacting directly with real-world websites. 
We select WebVoyager to evaluate the effectiveness of agents specifically designed for web-based tasks in retrieving and analyzing large-scale data from open domains. Since the web browser component of \sysname builds upon that of WebVoyager, it also serves as a baseline for comparing the design of the action space and reasoning mechanisms of \sysname.

\minihead{OpenAI Research Agent \cite{openai2024researchbot}} The OpenAI Research Agent is a recently released system introduced alongside OpenAI's new agent framework, which received more than 8.6K stars on GitHub within a month. It consists of a planner agent for generating search terms, a search agent for retrieving relevant information, and a report writing agent for composing answers. 
Since its search agent employs the same OpenAI web search tool used by the web augmenter in \sysname, it provides a natural point of comparison for in terms of the architectural design and coordination strategies of \sysname.

\minihead{OpenAI Web Search Tool \cite{openai2024tools} + TAG \cite{biswal2024text2sqlenoughunifyingai} (Search + TAG)} To explicitly demonstrate that \paradigmname extends beyond trivial integration of existing systems, we implement a baseline system composed of the OpenAI WebSearch Tool, representing the state of the art in \textbf{(C1)}, combined with the best-performing version of TAG using LOTUS, which supports both \textbf{(C2)} and \textbf{(C3)}. The search tool is executed for 10 iterations, with each iteration refining and extending the previous response, matching \sysname’s maximum number of iterations for a fair comparison.

\vspace{.5em}

All agents, including \sysname and the five baselines, are primarily powered by \texttt{gpt-4o-11-06}, with one explicit original design exceptions: the OpenAI Research Agent's writing agent uses \texttt{o3-mini} \cite{openai_o3_mini_docs_2024}. All agents are explicitly instructed not to access domains on the blacklist. For agents that support software-enforced blocking (e.g., via DNS), we apply such restrictions directly. Specifically, WebVoyager is prohibited from clicking links from blacklisted domains, and Google Custom Search API used by Deep Research is configured to exclude results from these domains.
Unless otherwise specified, agents follow the I/O requirements of \datasetname. The only exception is Search + TAG, which does not generate explicit code and therefore produces only outputs (1), (2), and (4) as defined in Section~\ref{subsec:task-desc}.

\subsubsection{Metrics} \label{subsec:metrics}

We apply two categories of metrics: system level and stage level, to quantitatively evaluate agent performance and cost efficiency on \datasetname. 

\minihead{System-level Metrics} 
We deploy accuracy as our primary performance metric, which measures whether the agent’s output (i.e., output (1) per definition in Section \ref{subsec:task-desc}) matches the ground-truth label. Any use of blacklisted domains is considered an automatic failure. To assess whether the agents' outputs are grounded in data retrieval and analysis, we evaluate the coherence among the retrieved data, the generated code, and the final result: specifically, whether the execution result of the generated code on the retrieved data aligns with the ground-truth label, namely the \textit{data-grounded (DG) accuracy}.
 For cost analysis, we report the average API costs for all requests made by each agent for each task.

\minihead{Stage-level Metrics}
We evaluate the performance of different \paradigmname stages by assessing the quality of their intermediate results.
Specifically, we examine the quality of the retrieved data (output of stages 1-2) and the generated code (output of stage 3). For data, we assess (1) data validity, i.e., whether the agent successfully collects data from open domains and organizes it into CSV format as a structured table, and (2) data similarity, specifically, schema similarity, with the ground-truth data. For code, we evaluate (1) the execution success rate, i.e., whether the code compiles, executes without error, and generates a result, and (2) code similarity with the ground-truth code.
Since there is no universally accepted gold standard for evaluating data and code similarity, we apply various evaluation methods to automate this process in a quantitative, comprehensive, and objective manner. We evaluate the performance of different \paradigmname stages by assessing the quality of their intermediate results. These similarity metrics are included primarily as diagnostic checks to better understand system behavior. 

\begin{table}[h]
\caption{Different similarity metrics applied to data and code.}
\centering
\resizebox{0.6\columnwidth}{!}{
\begin{tabular}{c!{\vrule width 0.8pt}cc}
\toprule
 & \textbf{LLM as a Judge} & \textbf{Embedding} \\
\toprule
\textbf{w/o Column Match }      &      $DataSim_1$      &     $DataSim_2$                    \\
\textbf{w/ Column Match}      &    $DataSim_3$                      &         $DataSim_4$                \\
\midrule
\textbf{w/o Normalization }     &       $CodeSim_1$      &      $CodeSim_2$                  \\
\textbf{w/ Normalization }    &       $CodeSim_3$       &                  $CodeSim_4$      \\
\bottomrule
\end{tabular}
}
\label{tab:evaluation_matrix}
\end{table}

We summarize the similarity metrics in Table \ref{tab:evaluation_matrix}. We use two backbone methods to evaluate similarity:
(1) LLM-as-a-judge: For data, we input \texttt{df.head()} for both the agent-retrieved and ground-truth data into \texttt{gpt-4o-mini} \cite{openai_gpt4o_mini_2024}, prompting it to generate a similarity score between 0 and 1.
For code, we follow the same procedure: both the agent-generated and ground-truth code are provided to the model to generate a similarity score.
(2) Embedding-based similarity: For data, we encode \texttt{df.head()} using \texttt{text-embedding-3-small} \cite{openai_embeddings_2024}, an embedding model that effectively captures tabular structure and content in text format. For code, we use CodeBERT \cite{feng2020codebertpretrainedmodelprogramming}, a model trained on large-scale code corpora and effective at capturing functional and structural similarities in code. We then compute the cosine similarity between the generated embeddings to obtain a numerical similarity score.

Given the nature of structured data and code, we apply alignment techniques to account for structural and semantic variability in similarity evaluation.
For data, since it consists of columnar structures, we perform \textit{column-level matching}. Each column is individually encoded using \texttt{text-embedding-3-small} \cite{openai_embeddings_2024}, and then matched to the most similar column in the corresponding ground-truth data based on cosine similarity of the embeddings.
For code, we implement a 3-stage custom normalization pipeline that standardizes both variable naming and symbolic expressions before measuring similarity: (1) Parse the code into Abstract Syntax Trees (ASTs) \cite{python_ast} (2) Rename user-defined variables to canonical forms (e.g., \texttt{var\_0}, \texttt{var\_1}, etc.), while preserving built-in names (3) Uses sympy to expand and simplify mathematical expressions in binary operations (e.g., both \texttt{x = (a + b) * (a + b)} and \texttt{x = (a + b) ** 2} are expanded to \texttt{a**2 + 2*a*b + b**2}) to preserve only the logical structure and semantics. This normalization is deterministic and converges to a unique normal form for a given input under the defined transformation rules.

\subsection{\sysname Achieves Promising Performance At Low Costs} \label{subsec:overall_evaluation}

\begin{table*}[h]
\caption{System-level performance and costs of different agents. 
Accuracy (Acc.) and Data-Grounded Accuracy (DG Acc.) are reported in percentages (\%), 
and cost is measured in \$0.01 units. 
The optimal value for each metric is highlighted with both bold and underline formatting.}
\centering
\resizebox{\textwidth}{!}{
\begin{tabular}{l|ccc|ccc|ccc}
\toprule
\multirow{2}{*}{\textbf{Agent}} 
& \multicolumn{3}{c|}{\textbf{Overall}} 
& \multicolumn{3}{c|}{\textbf{Verification}} 
& \multicolumn{3}{c}{\textbf{QA}} \\
\cmidrule(l){2-10}
& Acc.$\uparrow$ & DG Acc.$\uparrow$ & Cost$\downarrow$
& Acc.$\uparrow$ & DG Acc.$\uparrow$ & Cost$\downarrow$
& Acc.$\uparrow$ & DG Acc.$\uparrow$ & Cost$\downarrow$ \\
\midrule
Deep Research & 32.5  &  27.5  &   8.24    &  37    & 29      & 8.97 & 28  & 26 & 7.51 \\
AutoGPT &  21.5  &   4.0    &   9.36    &    29  &   6    & 9.53  & 14 & 2 & 9.19\\
WebVoyager &  46.5 &   20.5    &  5.29 &   49    &   16    & \textbf{\underline{6.33}} & 44 & 25 & 4.26 \\
OpenAI Research Agent &  12.5 &  12.5     &   33.94 & 12      &   12    &  33.95 & 13 & 13 & 33.93 \\
Search + TAG &  25     &  -     &   9.31    & 40      &   -    &  8.67 & 10 & - & 9.94\\
\midrule
\textbf{\sysname} & \textbf{\underline{86.5}}  &  \textbf{\underline{82.5}}   &   \textbf{\underline{5.03}}    &  \textbf{\underline{88}}     &  \textbf{\underline{80}}     &  7.62 & \textbf{\underline{85}} & \textbf{\underline{85}} & \textbf{\underline{2.45}}\\
\bottomrule
\end{tabular}
}
\label{tab:system_perf}
\end{table*}

\begin{table*}[h]
\caption{Accuracy (\%) of baseline agents without enforcing data and code generation under the \datasetname I/O format, compared with \sysname's performance. The optimal value for each metric is highlighted with both bold and underline formatting.}
\centering
\resizebox{0.9\textwidth}{!}{
\begin{tabular}{l|cccc|c}
\toprule
\textbf{Task} & Deep Research & AutoGPT & WebVoyager & OpenAI Research Agent & \textbf{\sysname} \\
\midrule
Overall & 19 & 17.5 & 28.5 & 7.5 & \textbf{\underline{86.5}} \\
Verification & 33 & 29 & 52 & 12 & \textbf{\underline{88}} \\
QA & 5 & 6 & 5 & 3 & \textbf{\underline{85}} \\
\bottomrule
\end{tabular}
}
\label{tab:no_io}
\end{table*}

As we can see from Table \ref{tab:system_perf}, \sysname achieves 86.5\% accuracy on all \datasetname tasks with an average cost of \$0.05, outperforming all the baseline agents across all metrics. We demonstrate our findings in detail as follows.

\minihead{\sysname significantly outperforms existing AI agents on \datasetname}  We observe that existing AI agents show limited capabilities in performing \datasetname tasks. Specifically, all baseline agents fail on more than half of the \datasetname tasks, with accuracies ranging from as low as 12.5\% to a maximum of only 46.6\%, making them unreliable for practical use. Notably, in verification tasks where the expected accuracy is 50\% through random guessing, none of the baseline agents reach this level. 

\sysname achieves over 85\% accuracy for both tasks, which is effective for practical deployment in real-world scenarios. 
It outperforms the baseline agents with an end-to-end accuracy of 2.7$\times$ as compared to Deep Research, 4$\times$ as compared to AutoGPT, a relative increase of 86\% for WebVoyager,
6.9$\times$ as compared to the OpenAI research agent, and 3.5$\times$ as compared to Search + TAG.

\minihead{\sysname reliably grounds answers in data} 
When it comes to DG accuracy, the performance of baseline agents further deteriorates, with the highest DG accuracy reaching only 27.5\%. For instance, while WebVoyager achieves a relatively high end-to-end accuracy of 46.5\%, its DG accuracy drops to 20.5\%, indicating that fewer than half of its correct answers are actually grounded in the retrieval and analysis of large-scale data. AutoGPT performs even worse, with only 4\% of its outputs being both correct and grounded, which is less than one-fifth of its original accuracy. These results highlight that existing agents often fail to generate reliable, data-grounded answers. Instead, they may rely on superficial information found on the web, treating it as truth without assessing source reliability or rigorous data analysis.

In contrast, \sysname achieves a DG accuracy of 82.5\%, indicating high reliability. Its architecture ensures that final results are derived from the execution of analytic code over retrieved data. The 4\% gap between its overall accuracy and DG accuracy reflects the fallback mechanism used when no relevant data is retrieved.

For the four baseline agents whose original outputs do not require generating data and code, we conducted ablated experiments following their native setups, where only the final result is required (i.e., they are not required to output data and code following the \datasetname I/O format). As we show in Table \ref{tab:no_io}, their performance remains largely unchanged on verification tasks but drops significantly on question answering tasks, as the lack of data grounding leads them to make unreliable guesses. This further underscores the importance of the data grounding principles introduced by \paradigmname.

\minihead{\sysname is cost-efficient in design} 
\sysname incurs the lowest cost among all baseline agents, demonstrating its lightweight and efficient design.

First, in terms of architecture, traditional general-purpose agents like AutoGPT integrate complex modules with extensive planning and coordination logic, leading to significant overhead. 
In contrast, \sysname adopts a streamlined architecture that retains only essential components required for data retrieval and analysis, resulting in an average cost that is nearly half that of AutoGPT.

Second, at the workflow level, existing agents tend to over-plan and execute redundant actions. For example, the OpenAI Research Agent generates an average of 8 search terms per task, resulting in a substantial cost overhead of 6.8$\times$. Similarly, Deep Research issues an average of 3 queries per task, incurring a cost of \$0.08, which is still notably higher than \sysname’s more efficient \$0.05.
In contrast, \sysname adopts a streamlined approach: it issues a single, focused query and follows through deeply. Although \sysname is capable of iterating through stages 1->2 in a multi-round fashion, it rarely needs to do so in practice: on average, only 1.41 rounds are required to retrieve the necessary information, highlighting the precision of its initial query formulation. 

Third, \sysname’s reasoning mechanism is optimized for cost efficiency. Even when compared to WebVoyager, which operates within a similar web-based action space, \sysname is more effective at extracting critical information. On average, \sysname performs 4.4 iterations of web browsing, while WebVoyager performs 2.29 iterations. 
Despite deeper explorations that lead to better performance, \sysname incurs lower costs due to its reasoning prompts, which are designed to guide the agent toward only the most essential information while minimizing unnecessary input and output tokens.

\minihead{\sysname addresses core deficiencies of existing baseline agents}
By examining detailed traces of baseline systems, we identify their key limitations and demonstrate how \sysname effectively addresses them:

Web search-based agents face challenges due to limitations in output diversity and length. Originally designed for simple point-lookups, these systems struggle with tasks requiring structured data manipulation. For example, WebVoyager can browse a webpage \cite{statista_minwage_states_2025} listing state minimum wages but fails to answer analytic questions such as ``How many states have minimum wages higher than the federal minimum wage in 2024?'' \cite{usafacts_minwage_2025}, which require operations like \texttt{COUNT} over structured data. Similarly, when tasked with identifying ``Which state has the highest rate of homelessness in 2024?'' \cite{usafactsHomelessness2024}, both Search+TAG and AutoGPT return incomplete results, unable to process and analyze full documents.
\sysname overcomes these limitations with an integrated interface that downloads, transforms, and analyzes data of diverse formats in a modular manner.

Research agents generate search queries, retrieve articles, and summarize them. However, they often rely on high-level textual summaries rather than verifying claims using original data sources, introducing an additional failure mode: hallucination from unreliable secondary sources. For instance, when evaluating the claim ``Uttar Pradesh is now the second-largest economy in the country, competing to become India's top economy'' \cite{indiantech2024up}, Deep Research references secondary articles that propagate misinformation. In contrast, \sysname grounds its reasoning in authoritative data \cite{statisticstimes_india_gdp}, reflecting the robustness of \paradigmname.

\minihead{\sysname produces interpretable and grounded end-to-end traces} 
Here we demonstrate how \sysname follows a coherent, multi-stage reasoning trace to verify a real-world factual claim.
In verifying the claim, ``The Federation for American Immigration Reform estimated that there are 16.8 million illegal aliens in the United States in 2023'' \cite{ramirezuribe2024politiFact}, \sysname first generates the search query \textit{``Federation for American Immigration Reform 16.8 million illegal aliens 2023''} and uses it to search on Google. It navigates to the official website of the Federation for American Immigration Reform \cite{fairus_organization}, locates the relevant report for 2023, and downloads the associated PDF file \cite{fairus_2023_illegal_alien_estimate}.

During the data transformation stage, \sysname identifies a plot on Page 1 of the PDF and transforms it into a structured CSV file with the schema:
\texttt{[Year, Estimated U.S. Illegal Alien Population (millions)]}.

For the analysis stage, \sysname generates a code snippet that retrieves the estimated population for the year 2023, semantically similar to the following SQL query: 
\begin{lstlisting}[language=CustomSQL]
SELECT "Estimated U.S. Illegal Alien Population (millions)"
FROM table WHERE Year = 2023
\end{lstlisting}

\subsection{\sysname Exhibits Reliable Stability} \label{subsec:generalizability}
We analyze the performance of \sysname and baseline agents across tasks with diverse features to assess their generalizability and stability.

\begin{table}[h]
\caption{The accuracy of different agents across tasks with different ground-truth labels and formats. T represents True, F represents False, N represents numeric values, and S represents strings. 
$\delta$ denotes the relative difference between two subtypes within the same task, and is calculated as $\delta = \frac{|\text{A} - \text{B}|}{\text{A} + \text{B}} \times 100$. The optimal value for each metric is highlighted with both bold and underline formatting.}
\centering
\renewcommand{\arraystretch}{1.2}
\resizebox{0.75\columnwidth}{!}{
\begin{tabular}{l|ccc|ccc}
\toprule
\multirow{2}{*}{\textbf{Agent}} & \multicolumn{3}{c|}{\textbf{Verification}} & \multicolumn{3}{c}{\textbf{QA}} \\
 & T (\%)$\uparrow$ & F (\%)$\uparrow$ & $\delta$ (\%)$\downarrow$ & N (\%)$\uparrow$ & S (\%)$\uparrow$ &$\delta$ (\%)$\downarrow$ \\
\midrule
Deep Research & 48 & 26 & 29.7  & 21.7 & 41.9 & 31.8 \\
AutoGPT & 2 & 56 & 93.1 & 8.7 & 25.8 & 49.6 \\
WebVoyager & 58 & 40 & 18.4 & 39.1 & 54.8 & 16.7 \\
OpenAI Research Agent & 10 & 14 & 16.7 & 14.5 & 9.7 & 19.8 \\
Search + TAG & 60 & 20 & 50 & 10.1 & 9.7 & \textbf{\underline{2.36}} \\
\midrule
\textbf{\sysname} & \textbf{\underline{84}} & \textbf{\underline{92}} & \textbf{\underline{4.5}} & \textbf{\underline{82.6}} & \textbf{\underline{90.3}} & 4.5 \\
\bottomrule
\end{tabular}
}
\label{tab:task_types}
\end{table}

\minihead{\sysname achieves stable performance across different task categories} 
Compared to verification tasks that require binary outputs, QA tasks demand more concrete answers, increasing the complexity of analytic reasoning involved. This added complexity is reflected in the accuracy drop observed in both \sysname and the baseline agents. As we show in Table \ref{tab:system_perf}, nonetheless, \sysname maintains strong and stable performance, achieving 85\% accuracy with only three fewer successful executions than in verification tasks. Notably, \sysname's data-grounded accuracy increases in QA tasks by 5 additional correct instances compared to verification, demonstrating its generalizability. In contrast, AutoGPT’s performance drops by half when transitioning from verification to QA tasks, indicating reduced adaptability to more complex tasks.

\minihead{\sysname generalizes across fine-grained task types within each category} 
As we show in Table~\ref{tab:task_types}, while baseline agents exhibit large disparities in performance across different task types, \sysname maintains consistently low variation, indicating strong generalization. Specifically, the relative difference in accuracy across all task types for \sysname remains below 5\%. Within verification tasks, the relative difference between accuracy on True and False instances for baseline agents ranges from 3.7 to 21.7 times higher than that of \sysname. In QA tasks, the disparity between performance on questions with numeric and string answers for baseline agents is 3.7 to 11 times greater than that of \sysname.

\begin{figure}[t!]
    \graphicspath{{figures/}}
    \centering
    \includegraphics[width=0.95\columnwidth]{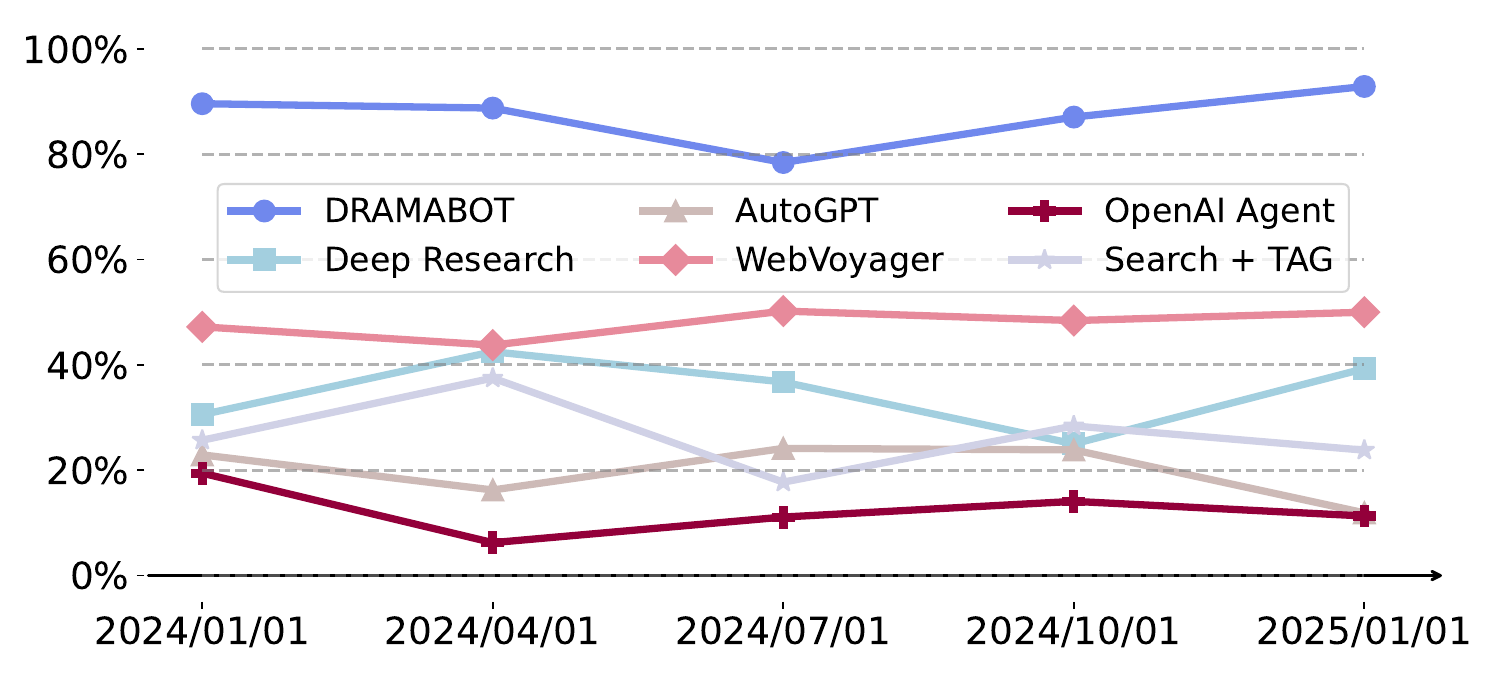}
    \caption{The overall accuracy (\%) of different agents across time. Each labeled time point marks the start of a three-month period.}
    \label{fig:across-time}
\end{figure}

\minihead{\sysname maintains superior performance across all time spans}  To evaluate performance over time, we assign each task a timestamp based on its publication date. Specifically, for Politifact claims, we use the original date the claim was made for the False claim and the date the corresponding claim was analyzed and posted on Politifact for the True claim; for Twitter/X Community Notes, we use the time of the original tweet for the False claim and the time when the corresponding community note was posted for the True claim; for USAFacts, we use the article publication date. 
We present the overall accuracy (in percentage) of different agents across task instances with varying timestamps in Figure~\ref{fig:across-time}. We can observe that the relative ranking of the baseline agents fluctuates over time. This indicates that \datasetname effectively captures the temporal dynamics of real-world data analysis tasks. Despite these fluctuations, \sysname consistently outperforms all baseline agents across all time periods, demonstrating its robustness to temporal shifts and its effectiveness even as the relevance and availability of information evolve over time.

\subsection{Stage-wise Strength Underpins \sysname's End-to-end Performance} \label{subsec:component_studies}

We evaluate \sysname’s stage-wise performance by assessing its intermediate outputs, specifically the quality of the retrieved data and the generated analytic code.
We summarize the quantitative results in Table \ref{tab:component_perf}, and demonstrate our findings in detail as follows.

\begin{table*}[t]
\caption{
Component-level performance of different agents. 
The optimal values of critical metrics are highlighted with both bold and underline formatting.
}
\centering
\renewcommand{\arraystretch}{1.2}

\subfloat[\textbf{Data Component}]{
\resizebox{\textwidth}{!}{
\begin{tabular}{l|c|cccc|c}
\toprule
\multirow{2}{*}{\textbf{Agents}} 
& \multirow{2}{*}{\begin{tabular}[c]{@{}c@{}}\textbf{Data Validity (\%)$\uparrow$}\end{tabular}} 
& \multicolumn{5}{c}{\textbf{Data Similarity}} \\
\cmidrule{3-7}
& & Sim1$\uparrow$ (rk$\downarrow$) & Sim2$\uparrow$ (rk$\downarrow$) & Sim3$\uparrow$ (rk$\downarrow$) & Sim4$\uparrow$ (rk$\downarrow$) & \textbf{\textit{Avg$\uparrow$ (rk$\downarrow$)}} \\
\midrule
Deep Research & 75.0 & 0.485 (4) & 0.461 (4) & 0.607 (4) & 0.435 (4) & \textit{0.497 (4)} \\
AutoGPT & 26.0 & 0.143 (6) & 0.134 (6) & 0.185 (6) & 0.125 (6) & \textit{0.147 (6)} \\
WebVoyager & 75.0 & 0.470 (5) & 0.433 (5) & 0.608 (3) & 0.427 (5) & \textit{0.484 (5)} \\
OpenAI Research Agent & 96.0 & 0.628 (2) & 0.571 (1) & 0.793 (2) & 0.530 (1) & \textit{0.631 (2)} \\
Search + TAG & 74.0 & 0.485 (3) & 0.464 (5) & 0.604 (2) & 0.440 (3) & \textit{0.498 (3)} \\
\midrule
\textbf{\sysname} & \textbf{\underline{97.5}} & 0.649 (1) & 0.557 (2) & 0.802 (1) & 0.525 (2) & \textbf{\underline{\textit{0.633 (1)}}} \\
\bottomrule
\end{tabular}
}
\label{tab:component_perf_data}
}

\vspace{1em}

\subfloat[\textbf{Code Component}]{
\resizebox{\textwidth}{!}{
\begin{tabular}{l|c|cccc|c}
\toprule
\multirow{2}{*}{\textbf{Agents}} 
& \multirow{2}{*}{\begin{tabular}[c]{@{}c@{}}\textbf{Code Execution (\%)$\uparrow$}\end{tabular}} 
& \multicolumn{5}{c}{\textbf{Code Similarity}} \\
\cmidrule{3-7}
& & Sim1$\uparrow$ (rk$\downarrow$) & Sim2$\uparrow$ (rk$\downarrow$) & Sim3$\uparrow$ (rk$\downarrow$) & Sim4$\uparrow$ (rk$\downarrow$) & \textbf{\textit{Avg$\uparrow$ (rk$\downarrow$)}} \\
\midrule
Deep Research & 61.5 & 0.420 (4) & 0.735 (4) & 0.406 (4) & 0.735 (4) & \textit{0.574 (4)} \\
AutoGPT & 11.0 & 0.328 (5) & 0.328 (5) & 0.258 (5) & 0.156 (5) & \textit{0.250 (5)} \\
WebVoyager & 66.5 & 0.667 (1) & 0.739 (3) & 0.673 (2) & 0.740 (3) & \textit{0.704 (3)} \\
OpenAI Research Agent & 94.5 & 0.611 (3) & 0.945 (1) & 0.684 (1) & 0.951 (1) & \textbf{\underline{\textit{0.798 (1)}}} \\
\midrule
\textbf{\sysname} & \textbf{\underline{95.0}} & 0.634 (2) & 0.932 (2) & 0.619 (3) & 0.937 (2) & \textit{0.781 (2)} \\
\bottomrule
\end{tabular}
}
\label{tab:component_perf_code}
}
\label{tab:component_perf}
\end{table*}

\minihead{Consistency across different similarity metrics supports robust comparison} We first evaluate the effectiveness of the similarity metrics by measuring their mutual agreement rate. This rate is defined as the average pairwise agreement across all metrics, where each pair determines whether the relative ranking of items is preserved. A higher mutual agreement rate indicates stronger consistency among the metrics in capturing item-level similarity.

For data similarity, the mutual agreement rate is 82.2\%. All four metrics consistently reflect the same overall trend: \sysname and the OpenAI Research Agent rank highest with similar scores, followed by Deep Research, Search + TAG, and WebVoyager, with AutoGPT consistently ranked lowest.
For code similarity, the mutual agreement rate is 83.3\%, with the metrics agreeing on the relative rankings: the OpenAI Research Agent produces the most similar code, followed by \sysname, WebVoyager, Deep Research, and AutoGPT.

\minihead{Strong correlation across \paradigmname stages highlights the benefit of unification} 
The quality of the collected data and the generated code are highly correlated, with a Pearson coefficient \cite{pearson1895regression} of $\rho = 0.97$ between the average data similarity and code similarity. This highlights the significance of \paradigmname in unifying the pipeline to ensure consistency and effectiveness across both data retrieval and code generation.

\minihead{\sysname retrieves data of high quality}
We observe that \sysname retrieves data with the highest validity rate, reaching up to 3.8$\times$ of that of AutoGPT. It also achieves the highest average data similarity, up to 4.3$\times$ of that of AutoGPT. These results demonstrate the effectiveness of the data retriever in both the data collection and data transformation stages in \paradigmname.

\minihead{\sysname's data extraction achieves 90\% accuracy}
We rerun \sysname on the 27 failure cases by providing direct access to the collected data, either the original websites where the data is embedded or the readily downloaded raw files.
 Since successful end-to-end task completion indicates accurate data extraction, \sysname achieves 90\% extraction accuracy across all \datasetname tasks, with 92\% for verification tasks and 88\% for QA tasks. This demonstrates the effectiveness of \sysname's strategy.

\minihead{\sysname demonstrates strong analytic reasoning capabilities}
We observe that \sysname achieves the highest code execution rate, reaching up to 8.6$\times$ that of AutoGPT. Although \sysname ranks second in code similarity, the difference from the highest score is less than 0.02. 

\subsection{Web Browser and Web Augmenter Coordinate for Data Collection} 
\label{subsec:ablation_studies}

We proceed to analyze the behavior of the data retriever to understand its effectiveness in detail. We first examine how it coordinates the web browser and the web augmenter, which together execute the data collection stage in \paradigmname, and then evaluate their individual contributions.

\begin{table}[h]
\caption{The percentage of task instances where the data is collected by different subagents of \sysname and their corresponding accuracy.}
\centering
\renewcommand{\arraystretch}{1.2}
\resizebox{0.58\columnwidth}{!}{
\begin{tabular}{l|c|c}
\toprule
\textbf{Sub-agent} & \textbf{Workload Share (\%)} & \textbf{Accuracy (\%)} \\
\midrule
Web Browser    & 63.5 & 91.2 \\
Web Augmenter  & 36.5 & 68.0 \\
\bottomrule
\end{tabular}
}
\label{tab:browser-augmenter}
\end{table}

\minihead{Data retriever successfully coordinates the web browser and web augmenter for balanced workload and accuracy} We can see from Table~\ref{tab:browser-augmenter} that in 63.5\% of the task instances, the web browser collects the data, while the remaining 36.5\% are handled by the web augmenter. This demonstrates that the coordinating capability of the data retriever results in a reasonable division of labor that aligns with each sub-agent’s accuracy. Specifically, the web browser, which explores data at a finer level of granularity and is more effective at handling \datasetname tasks, takes on a larger share of the workload to ensure strong overall system performance.

\begin{table}[t]
\caption{Accuracy (\%) of \sysname on different \datasetname tasks with different components removed.}
\centering
\renewcommand{\arraystretch}{1.2}
\resizebox{0.85\columnwidth}{!}{
\begin{tabular}{l|c|c|c|c}
\toprule
\textbf{Setting} & \textbf{Overall} & \textbf{Verification} & \textbf{QA} & \textbf{Workload Share (\%) in Table \ref{tab:browser-augmenter}} \\
\midrule
Web Browser only & 59 & 57 & 61 & 60.7 \\
Web Augmenter only & 53 & 59 & 47 & 52.1 \\
\bottomrule
\end{tabular}
}
\label{tab:ablation}
\end{table}

\vspace{.5em}
We further conduct ablation analysis by removing the web browser and the web augmenter individually, and evaluate the system on all \datasetname tasks.

\minihead{Both web augmenter and web browser enhance \sysname's performance} 
As we compare the performance of \sysname with different sub-agents removed in Table~\ref{tab:ablation} to that of the original \sysname in Table~\ref{tab:system_perf}, we observe that both subagents play a critical role in ensuring the system's overall effectiveness. Specifically, removing the web browser results in a 33.5 percentage point drop in performance, while removing the web augmenter leads to a 27.5 point drop.

\minihead{Both web augmenter and web browser demonstrate strong standalone effectiveness} We observe in Table~\ref{tab:ablation} that although the accuracy of \sysname decreases when individual sub-agents are removed, both the web augmenter and the web browser alone still deliver strong performance in data retrieval and coordinate well with the remaining components, outperforming all baseline agents as we present in Table~\ref{tab:system_perf}. Specifically, WebVoyager, which serves as the foundation for \sysname's web browser, has an overall accuracy that is 12.5 percentage points lower than the version of \sysname using only the web browser. Similarly, the OpenAI Research Agent, which also relies on the OpenAI web search tool for data collection, achieves less than 1/4 the accuracy of the version of \sysname using only the web augmenter. These results highlight two key findings: (1) the overall \sysname architecture effectively and reliably orchestrates its components to maintain high performance, even when certain modules are removed or operating suboptimally; and (2) the reasoning mechanisms adapted to the individual sub-agents are robust and capable of maintaining strong task performance in isolation.

\minihead{The interaction between web augmenter and web browser improves each other's performance}
As we compare the performance of \sysname with a single sub-agent in Table~\ref{tab:ablation} to the case where the full \sysname executes and collects data using that same sub-agent in Table~\ref{tab:browser-augmenter}, we observe that the interaction between the web browser and the web augmenter enhances each other’s performance.
Specifically, when the web browser collects data in the presence of the web augmenter, its accuracy improves by 30.5 percentage points compared to operating alone. Similarly, when the web augmenter collects data with support from the web browser, its accuracy increases by 15 percentage points. This indicates that the integration of the web browser and the web augmenter goes beyond functioning as a simple oracle. Instead, it reflects a synergistic relationship among \sysname’s sub-agents, leading to mutually reinforced improvements in data retrieval quality and reasoning capabilities.
\section{Related Work}\label{sec:related_work}
We review related literature in the following three areas.

\minihead{Fact-checking models and benchmarks} 
Existing fact-checking benchmarks \cite{tang2024minicheckefficientfactcheckingllms, tang2024tofuevalevaluatinghallucinationsllms, kamoi-etal-2023-wice, liu-etal-2023-evaluating, wang2024factcheckbenchfinegrainedevaluationbenchmark} primarily focus on text retrieval or statement-level summarization, without requiring precise answers grounded in analytic reasoning. For example, ClaimVerify \cite{liu-etal-2023-evaluating} verifies claims such as "What are the latest discoveries from the James Webb Space Telescope?" using binary factual consistency annotations for each cite-worthy sentence. This setup centers on information extraction rather than performing grounded analysis or calculations.
These limitations highlight the gap in existing benchmarks, which are insufficient for evaluating the full \paradigmname pipeline, and underscores the significance of \datasetname in advancing benchmarks for analytic, data-grounded fact verification.

Existing fact-checking models fall short in important ways for end-to-end data discoveries. Models that rely on web search \cite{10.1145/988672.988687, 10.5555/2283396.2283398, kamp2023openinformationextractionreview} often lack the ability to process long and complex information locally, such as reasoning over full documents. On the other hand, models that operate over local documents \cite{tang2024minicheckefficientfactcheckingllms, kryscinski-etal-2020-evaluating, fabbri-etal-2022-qafacteval, goyal-durrett-2021-annotating, laban, min-etal-2023-factscore, weng-etal-2023-large, gao-etal-2023-rarr, malaviya2024expertqaexpertcuratedquestionsattributed}, including those based on the RAG paradigm \cite{lewis2021retrievalaugmentedgenerationknowledgeintensivenlp, Chen_2023}, assume access to a local, static, and predefined set of data, and are therefore unable to retrieve information from open-domain sources.
These limitations highlight the need for a unified approach that integrates both data collection and data transformation for more effective fact verification. This is precisely the problem that \paradigmname is designed to solve, and \sysname demonstrates its capability to carry out this unified pipeline in practice.

\minihead{Dynamic and up-to-date knowledgebases} Knowledge bases have significantly enhanced the capabilities of LLMs in a variety of knowledge-intensive tasks \cite{wang2023knowledgptenhancinglargelanguage}. Although there has been substantial progress in constructing up-to-date knowledge bases \cite{liška2022streamingqabenchmarkadaptationnew, kasai2024realtimeqawhatsanswer} to keep models aligned with recent information and to evaluate their ability to identify and extract time-sensitive knowledge, these updates are performed manually. The models themselves do not actively retrieve new information.
For example, RealTime QA \cite{kasai2024realtimeqawhatsanswer} provides a regularly updated evaluation platform, with updates released weekly in the current version. However, the retrieval process is manually configured, limited to a fixed dataset, and not based on open-domain access. Most importantly, from the perspective of the model, the data is treated as local, static, and predefined. Streaming QA represents a similar case, where data from all dates is made available at once rather than continuously updated \cite{liška2022streamingqabenchmarkadaptationnew}.
These limitations underscore the need for systems that support dynamic, open-domain information retrieval and reasoning. This is precisely the challenge that \paradigmname is designed to address.

\minihead{Intelligence data systems} The development of large language models (LLMs) has enabled the automation of many tedious aspects of data analysis, ranging from data preprocessing \cite{Hu_2024, cao2024spider2vfarmultimodalagents, hong2024datainterpreterllmagent} to semantic querying \cite{biswal2024text2sqlenoughunifyingai, zhang2024datacopilotbridgingbillionsdata, dail_sql, patel2025semanticoperatorsdeclarativemodel, liu2024declarativeoptimizingaiworkloads, shankar2025docetlagenticqueryrewriting, eisenschlos-etal-2020-understanding, hu-etal-2025-repro}. However, all of these systems operate over a static collection of data. With the exception of TAG \cite{biswal2024text2sqlenoughunifyingai}, which integrates retrieval-augmented generation with NL2SQL to construct a query-driven database for more efficient information retrieval and execution, existing systems assume that a clean and well-formatted database is already available at query time.
This strong assumption limits their applicability to real-world tasks, where data must often be collected, transformed, and queried dynamically. In contrast, \paradigmname addresses this gap by enabling an end-to-end workflow that combines open-domain data collection with flexible, query-driven analysis.
\section{Conclusion}
In this paper, we propose \paradigmname, an end-to-end paradigm that automates data science workflows by unifying data collection, transformation, and analysis. We further collect \datasetname, a benchmark consisting of 200 task instances that require retrieving and analyzing data from open domains. 
Alongside \datasetname, we develop \sysname, a multi-agent system composed of a data retriever and a data analyzer. 
By coordinating the execution of sub-agents and optimizing the reasoning mechanisms, \sysname achieves 86.5\% accuracy on \datasetname at an average cost of only \$0.05 per task, significantly outperforming all five state-of-the-art baseline agents. 

\begin{acks}
We thank Jeonghwan Kim for his valuable insights on RAG systems; Qiusi Zhan, Yuxuan Zhu, and Tengjun Jin for their helpful advice on paper writing; and Yurong Liu and Zishan Su for their generous support during the revision process.
\end{acks}

\bibliographystyle{ACM-Reference-Format}
\bibliography{reference}

\end{document}